\documentclass[a4paper,11pt]{article}
\usepackage{pos}
\usepackage[prefix, nopostdot, nonumberlist]{glossaries-extra}
\usepackage{xspace}
\usepackage{lineno}
\usepackage{svg}
\usepackage{lipsum}
\usepackage{setspace}
\usepackage{natbib}
\usepackage{enumitem}
\usepackage{wrapfig}

\renewcommand{\glossarysection}[2][]{}
\newcommand{\mynewabbreviation}[4][]{%
  \newabbreviation[#1]{#2}{#3}{#4}%
  \expandafter\newcommand\csname #2\endcsname{\gls{#2}\xspace}%
  \expandafter\newcommand\csname #2s\endcsname{\glspl{#2}\xspace}%
  \expandafter\newcommand\csname a#2\endcsname{\pgls{#2}\xspace}%
  \expandafter\newcommand\csname an#2\endcsname{\pgls{#2}\xspace}%
}
\setabbreviationstyle[long]{long}
\glsdisablehyper
\GlsXtrEnableEntryCounting{abbreviation}{1}
\input{acronyms.acr}

\newcommand{\degree}{$^\circ$\xspace}
\newcommand{\dmax}{d_\text{max}}
\newcommand{\fluence}{\mathscr{F}}
\newcommand{\vxB}{{\vec v \times \vec B}}

\newcommand{\Offline}{\mbox{$\overline{\textrm{Off}}$\hspace{.05em}%
    \raisebox{.4ex}{$\underline{\textrm{line}}$}}\xspace}

\title{Event reconstruction with the Radio detector of the Pierre Auger Observatory}

\author*[a]{Simon Str\"ahnz}

\onbehalf{for the Pierre Auger collaboration$^b$}

\affiliation[a]{Karlsruhe Institute of Technology,\\
  Kaiserstraße 12, Karlsruhe, Germany}
\affiliation[b]{Observatorio Pierre Auger, Av.\ San Mart{\'\i}n Norte 304, 5613 Malarg\"ue, Argentina\\
Full author list: {\rm\url{https://www.auger.org/archive/authors_icrc_2025.html}}}

\emailAdd{spokespersons@auger.org}

\abstract{
The surface detector of the Pierre Auger Observatory has recently been
upgraded with the addition of radio antennas, forming the \rd.
This contribution outlines the standard methods for reconstructing extensive air
showers using the \rd, along with recent developments.

The reconstruction pipeline is based on a robust understanding of the detector
itself. The entire instrument, including the antenna pattern and analog chain,
has been meticulously characterized within the \Offline software framework,
based on measurements in the laboratory as well as in the field.
To ensure data integrity, stations identified as unreliable through monitoring
are excluded before event reconstruction. Absolute calibration is achieved at
the 5\% level by analyzing the diffuse galactic radio emission. Next, the
electric field that induced voltages in the antenna is calculated by
``unfolding'' the antenna response pattern. Key observables, such as the energy
fluence (the energy deposited in the ground per unit area) and the arrival time
of the pulse, are then determined. With these quantities, shower parameters can
be reconstructed with very good accuracy in two $\chi^2$-minimization fits: one
to determine the shower’s arrival direction via a spherical wavefront fit
(predicted within $0.2^\circ$), and the other to estimate the distance to the
shower maximum and the electromagnetic cascade energy (predicted within 5\%) using a lateral density
function.
}

\ConferenceLogo{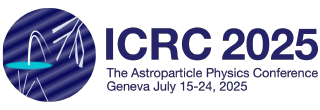}

\FullConference{%
39th International Cosmic Ray Conference (ICRC2025)\\
15 -- 24 July, 2025\\
Geneva, Switzerland}

\begin{document}
\maketitle

\setlength{\parskip}{0cm}
\setlength{\itemsep}{0cm}
\setlength{\bibsep}{0cm}

\section{Introduction}
    The Pierre Auger Observatory is currently the largest observatory for cosmic rays in the world.
    Situated in the province of Mendoza in Argentina, it covers an area of $\sim3000$\,km$^2$ with $\sim1600$ \sd stations with \wcds in a triangular 1.5\,km grid, overlooked by 27 fluorescence telescopes in 4 sites.
    Recently, the \sd stations have been upgraded with the addition of scintillators and radio antennas.
    The latter measure \eass by detecting the coherent radio emission generated due to the acceleration of charged particles (mainly electrons and positrons) during the shower development.
    For now, the readout of the antennas is triggered by the \wcds, but further triggering possibilities are under investigation.
    Two emission processes can be described macroscopically~\cite{huegeRadio}: The dominant emission process is due to charged particles being deflected in the geomagnetic field and is thusly called ``geomagnetic emission''.
    The other process is called ``charge-excess'' (CE) emission. It arises due to the separation of negative charges in the shower front from positive charges in its wake. 

    The next sections will explain the steps necessary to reconstruct an event from the measured data in the form of ``time traces'' (measured voltage over time).
    To perform a good reconstruction it is necessary to prepare the measured traces by checking the calibration of the measurement equipment and cleaning the traces from narrow-band \rfi.
    The reconstruction then starts with extracting the cosmic ray signal from the measured voltages.
    The timing of the signal can then be fitted to the expected shape of the wavefront to reconstruct the arrival direction.
    A second fit to the magnitudes of the signal can be used to find the energy in the electromagnetic cascade and the distance to the shower maximum, albeit with limited resolution for the latter.
    All necessary steps, as well as a full description of the \rd hardware, have been implemented within the \Offline software framework, which is the standard for the observatory.

\section{Preparation}
    \subsection{Calibration}
    \begin{figure}
        \centering
        \includesvg[width=0.65\linewidth]{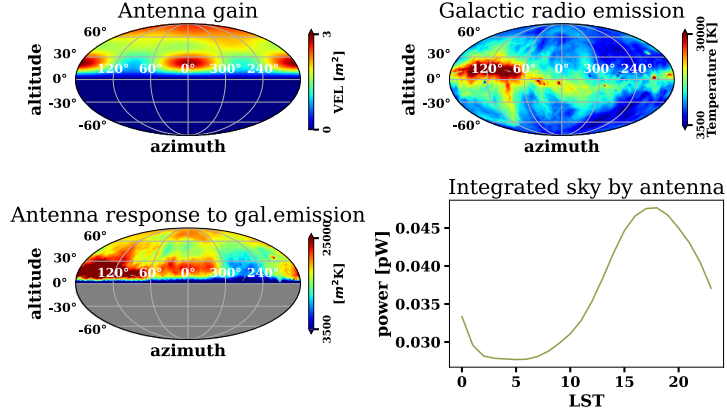}
        \caption{Example of the calculation of the antenna response to a diffuse radio emission model in local coordinates. \emph{Top left:} The antenna gain pattern. \emph{Top right:} The emission model. \emph{Bottom left:} The combination of the antenna response and the galactic emission. \emph{Bottom right:} The integral of the sky as function of local siderial time, which is used as calibration reference. Figure from~\cite{fodran_2024_thesis}}
        \label{fig:galcal}
    \end{figure}
    To verify the absolute calibration of the antennas in the field, the diffuse galactic radio emission can be used~\cite{fodran_2024_thesis}.
    For cosmic ray measurements, the galactic emission is just noise, but since it is relatively well known \cite{Busken-paper-GalCal} and present in all measurements, it can be used as a standard candle to achieve absolute calibration. 
    
    There are two different ways to obtain datasets for this calibration, both of them have been used to establish this method of calibration.
    The first is to extract traces from \CR measurements, as not all stations that are read out in an event will have a cosmic ray signal.
    To find those stations without signal, cuts on the maximum value and the standard deviation of the recorded time trace have been used. The advantage of this method is that it incorporates data from multiple positions within the array.
    The second method is to go into the field and record time traces directly from individual stations, which has the advantage of more control about the time of recording but less overall coverage of the array.

    To compare those measurements to the galactic emission on voltage level, the antenna response to that emission must be calculated.
    The first complication to that end is that the exact emission is not entirely known. There is a multitude of models for the galactic radio emission, each with their own advantages and disadvantages.
    Since it is not clear which model offers the best (or even just better) description, calibration constants are calculated separately for each and averaged.
    These calculations include the antenna response and the response of the analog chain, which must be incorporated (figure~\ref{fig:galcal}, top left, to right, to bottom left).
    To find the total power as a function of \lst, the emission must be integrated over the whole visible sky (figure~\ref{fig:galcal}, bottom right).
    The average calibration constants agree with measurements of the RD signal chain and simulated antenna gain patterns in the lab within 3\%, well within the uncertainty introduced by the variation between different sky models.
    
    Efforts have also been made to verify the simulated antenna response pattern in situ with drone-based measurements, which are still ongoing~\cite{reuzki_2024_arena}.
    
    \subsection{Cleaning}
    
    The measured signal will contain narrowband \rfi for two main reasons:
    First, there are the radio beacon transmitters deliberately installed by the observatory, which are used for the high precision timing calibration necessary for interferometric studies.
    Second, there is a significant number of unrelated transmitters, as the observatory was not built in a radio quiet zone.
    To remove these noise sources a filter is implemented based on the \dtft, which is defined by
    \begin{equation}
        \tilde F(f) = \sum_{n=0}^N F_n e^{-2\pi i ft_n},
    \end{equation}
    where $\tilde F(f)$ is the frequency-dependent transform of the discrete $F_n$ at times $t_n$.
    In contrast to the \dft often used in signal processing, the \dtft is continuous\footnote{
        For calculations, the frequency must of course also be discretised, but with a \dtft this frequency binning is not tied to the time binning as it is in a \dft. In this case, a frequency binning 20 times finer than the resolution of the \dft is used.
    } in frequency space.
    Given a discretely sampled time series of measurements, the \dtft can determine the amplitude and phase of any monochromatic contribution to the measurement with high precision and is not limited to the frequency-resolution of the \dft (see figure~\ref{fig:dtft}).
    This property can be used to identify and remove single-frequency transmitters using the following steps:
    \begin{enumerate}
        \setlength{\itemsep}{0cm}
        \setlength{\parskip}{0cm}
        \item Apply a \dft to the measured time series and identify the frequency bin with the highest amplitude
        \item Apply a \dtft within that bin and find the frequency, amplitude, and phase of the transmitter
        \item Subtract the corresponding waveform from the time series
    \end{enumerate}
    This is repeated up to 5 times while the relative power to be removed remains larger than 4\%.
    An example of the effect of this cleaning algorithm can be seen in figure~\ref{fig:dtft_example}.
    \begin{figure}
        \centering
        \begin{minipage}{0.45\linewidth}
            \centering
            \includesvg[width=\linewidth]{DTFT.svg}
            \caption{Spectrum of a cosine wave  analysed with a \dft (blue dots) and \dtft (green line). The frequency of the wave is shown by the orange line.}
            \label{fig:dtft}
        \end{minipage}\hspace{1cm}%
        \begin{minipage}{0.45\linewidth}
            \centering
            \vspace{-0.4cm}
            \includesvg[width=\linewidth]{spectrum_comparison.svg}
            \caption{Example spectrum of an \eas signal before (green) and after (blue) being cleaned with the \dtft method}
            \label{fig:dtft_example}
        \end{minipage}
    \end{figure}

\section{Reconstruction}
    \subsection{Signal}
    The radio signal, for the sake of this reconstruction method, is defined by the energy fluence (the energy deposited per unit area) and the peak time of the radio pulse emitted during the development of the air shower. 
    Both are determined from the reconstructed electric field.
    
        \subsubsection*{Electric Field}
        The electric field is calculated by ``unfolding'' the antenna and analogue response from the measured voltages.
        The voltage $V_i$ in channel $i$, as measured by the \adc, is related to the electric field ($\vec E$) in the frequency domain by
        \begin{equation}\label{eq:E_to_V}
            V_i = R_i\vec{H_i}(\theta, \varphi)\cdot\vec{E},
        \end{equation}
        where $\vec H_i$ is the \vel matrix describing the antenna response pattern and $R_i$ is the analogue response of channel $i$.
        The response of the analogue chain was measured in the laboratory.
        The \vel was determined by simulations made with 4NEC2~\cite{nec,giaccari_2022_antennapattern} which are currently being confirmed and possibly refined by \theDword-based measurements in the field~\cite{reuzki_2024_arena}.

        Inverting equation~\eqref{eq:E_to_V} to calculate the electric field from the measured voltage is not entirely straightforward.
        Since the \rd only measures two polarisations, it is necessary to assume the far field approximation for transverse waves, i.e.\ that the plane of polarisation is perpendicular to the direction of propagation ($E_r = 0$).
        In a first step, to solve the resulting 2-dimensional linear system of equations, the \vel matrix is checked for values close to zero, which simplify one equation, so that it can be solved directly.
        Otherwise, the system is solved using Cramer's rule.
        In the case that the determinant of the \vel matrix is too close to zero for both permutations of the equations, the system is deemed uninvertible, and it is assumed, that the corresponding electric field is zero.
        Lastly, the electric field is upsampled by a factor of 4 to improve the resolution of the peak finding and integration in the next step.

        \subsubsection*{Arrival Time and Energy Fluence}
        To find the arrival time, first a Hilbert envelope is applied to the electric field.
        Then the maximum bin within a predefined signal window is taken as the arrival time of the signal.
        For the uncertainty of the arrival time, 3~contributions must be considered:
        The uncertainty due to finite binning, the uncertainty introduced by noise, and the uncertainty of the GPS clock.
        The finite binning introduces an uncertainty of $\delta_t/\sqrt{12}$, where $\delta_t$ is the bin width.
        A simulation study~\cite{glaser_phd_thesis} has shown that the uncertainty due to noise depends on the signal-to-noise ratio\footnote{
            The signal to noise ratio is defined as: $\mathrm{SNR} = (|E|_\text{max} / |E|_\text{RMS} )^2$
        } and can be described by $\sigma_\text{noise} = 11.7\,\text{ns}\ \mathrm{SNR}^{-0.71}$. The GPS accuracy is taken from the detector description.
        For the current generation of hardware it is 5\,ns, although a future Auger-wide beacon system can reduce this to 1\,ns. These contributions are added in quadrature.
        
\begin{figure}
            \centering
            \includesvg[width=\linewidth]{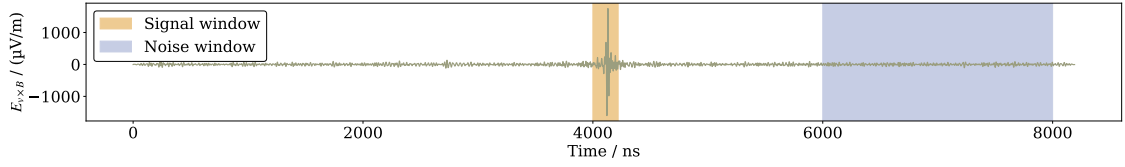}
            \caption{Example of a time trace measured with the \rd. The area highlited in orange is the signal window, the area in blue is the noise window.}
            \label{fig:trace_example}
        \end{figure}

        The energy fluence~($\fluence$) can be calculated by integrating the Poynting flux over time.
        For transverse waves, this simplifies to $\fluence = \epsilon_0 c \int |\vec E|^2 \mathrm{d}t$.
        To find the energy fluence corresponding to the emission from the air shower, background noise must be taken into account.
        This is done using the so called ``noise subtraction'' method:
        In this method, it is assumed, that the measured signal is the superposition of the signal field~($S$) and a noise field ($\vec E(t) = \vec S(t) + \vec e(t)$), whereby the noise~($\vec{e}$) is assumed to follow a normal distribution with $\mu = 0$ and $\sigma = \sigma_e$.
        The time integral is calculated separately in a narrow signal window (between $t_1$ and $t_2$, 220\,ns), which is centred at the signal time, and in a wider noise window (between $t_3$ and $t_4$, 2\,µs) at the edge of the recorded time series (see also figure~\ref{fig:trace_example}).
        Assuming, that the noise power is constant over the length of the recorded time series ($\sim$8\,µs), the integrated power in the noise window can be rescaled to the size of the signal window and then subtracted:
        \begin{equation}
            \fluence = \epsilon_0 c \int_T |\vec S|^2 \mathrm{d}t = \epsilon_0 c \left(\int_{t_1}^{t_2}|\vec E|^2\mathrm{d}t - \frac{t_2-t_1}{t_4-t_3}\int_{t_3}^{t_4}|\vec E|^2\mathrm{d}t\right)
        \end{equation}
        From the assumption of superposition it follows that the uncertainty of the fluence is given by
            $\sigma^2_{S+e} = 2\epsilon_0c\delta_t (2\fluence\sigma^2_e + \epsilon_0c(t_2-t_1)\sigma^4_e)$~\cite{glaser_phd_thesis} .
        The gain uncertainty of each individual antenna~($\sigma_\text{G}$) and between antennas~($\sigma_\text{A-A}$) must be added in quadrature. Since this model does not produce sufficiently large uncertainties for small signals, the energy Fluence of the noise window ($\fluence_N$) is added to the uncertainty. So the total uncertainty is\footnote{
            The factors 2 are due to the quadratic relationship between fluence and amplitude
        }
            $\sigma^2_\fluence = \sigma^2_{S+e} + (2\fluence\sigma_\text{G})^2 + (2\fluence\sigma_\text{A-A})^2 + \fluence_N^2$.
    
    \subsection{Direction}
\begin{figure}
        \centering
        \includesvg[width=0.5\linewidth]{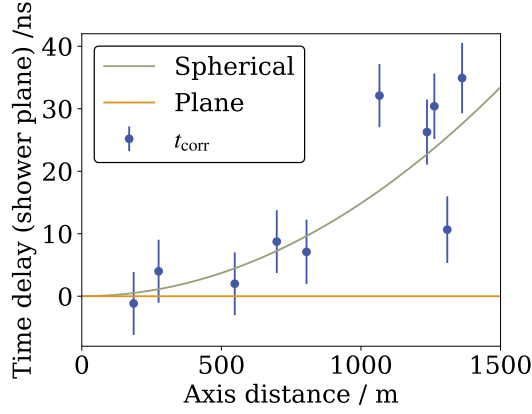}
        \caption{Example of the measured arrivals times in shower plane (i.e. with the time delay coresponding to a plane wave removed). For a plane wave the arrival times would match the yellow line at zero. For reconstruction, a spherical wave front (green line) is assumed.}
        \label{fig:wavefronts}
    \end{figure}

    Directional reconstruction is achieved by fitting the arrival time of the radio emission at each station to the expected shape of the wavefront using $\chi^2$ minimisation.
    There have been extensive studies of the shape of the wavefront based on measurements and simulations.
    Studies by LOFAR and LOPES show that a hyperbolic wavefront offers the best fit to measurements of vertical air showers \cite{lofarWavefront, lopesWavefront}.
    A study based on a library of CoREAS-simulated \cite{huegeCoREAS} air showers with 10~EeV proton and iron nuclei as primary and zenith angles ranging from 60\degree to 85\degree has shown, that for the situation at the Pierre Auger observatory, a spherical wavefront is sufficient to describe inclined air showers~\cite{gottowik_phd}. This can also be seen in the example in figure~\ref{fig:wavefronts}.

    \subsection{Energy and Distance to Shower Maximum}
    Energy and an estimate of the distance to shower maximum can be reconstructed by fitting the lateral distribution of the energy fluence.
    In general, this \ldf is 2-dimensional, which can be understood by the superposition of the two macroscopic emission processes mentioned above.
    These two processes emit radiation with different planes of polarisation, depending on the relative position of the observer to the shower axis.
    Another effect that is dependent on the observer position is the ``geometric early-late'' effect. For inclined showers, an observer placed towards the arrival direction is significantly closer to the emission region than an observer placed in the opposite direction.
    Combined, these effects produce a lateral distribution that is not rotationally symmetric.
    
\begin{figure}
        \centering
        \includesvg[width=0.5\linewidth]{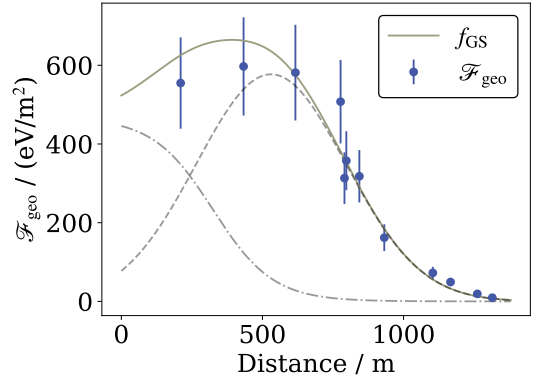}
        \caption{Example of the \ldf fit from an air shower. The solid line shows the fit function (eq.~\ref{eq:gs}), the dashed and dash-dotted line the Gaussian and the sigmoid respectively. The blue dots do not show the measured $\vxB$ fluence, which is used in the fit, but rather the geomagnetic fluence calculated from that. The $\vxB$ fluence cannot be described by a one-dimensional function (see eq.~\ref{eq:geo_vxb}).}
        \label{fig:ldf_example}
    \end{figure}

    For this reconstruction a signal model has been developed~\cite{schlueter_2023_signal_model} that makes use of the well known dependencies of these effects to formulate a 1-dimensional \ldf.
    The early-late effect can be addressed by a correction factor
        $c_\text{el} = 1 + \frac{z_i}{\dmax}$
    (with $z_i$ the distance between observer $i$ and the shower plane and $\dmax$ the distance to shower maximum) and the respective corrections on energy fluence and axis distance
        $\fluence = c_\text{el}^2\fluence_\text{raw},\ r = \frac{r_\text{raw}}{c_\text{el}}.$
    The fluence from geomagnetic emission alone, if corrected in this way, is rotationally symmetric w.r.t.\ the corrected axis distance.
    The lateral distribution of the geomagnetic emission can then be described by this function $f_\text{GS}$:
    \begin{equation}\label{eq:gs}
        f_\text{GS} = f_0 \left[\exp\left(-\left(\frac{r-r_0^\text{fit}}{\sigma}\right)^{p(r)}\right) + \frac{a_\text{rel}}{1+\exp(s [r/r_0^\text{fit} - r_{02}])} \right]
    \end{equation}
    The first half of this function is a Gaussian, defined by its position~$r_0^\text{fit}$, size~$\sigma$ and exponent~$p(r)$, the second half is a sigmoid with amplitude~$a_\text{rel}$ relative to the Gaussian, position~$r_{02}$ and size~$s$.
    The overall normalisation is set by parameter~$f_0$.
    The shape parameters ($r_0$, $p(r)$, $\sigma$, $a_\text{rel}$, $r_{02}$ and~$s$) have been parametrised by the shower observable $\dmax$ (the distance to the shower maximum) (see Appendix C in~\cite{schlueter_2023_signal_model}).
    The overall normalisation depends on the observable $E_\text{geo}$ (the energy radiated in the form of geomagnetic emission), the distance to the shower axis on the core position $\vec x$.
    These observables are then used as parameters for the fit.
    An example of this can be seen in figure~\ref{fig:ldf_example}.

    Up to this point the model only describes the energy fluence due to geomagnetic emissions, which, although the dominant emission mechanism, cannot be directly measured.
    To model a measureable quantity (usually the energy fluence in the $\vxB$ direction, where $\vec v$ is the propagation direction of the shower and $\vec B$ the geomagnetic field) it is necessary to also take the \ce emission into account.
    For this, the \ce fraction $a_\text{ce} \equiv \sin^2(\alpha) \fluence_\text{ce}/\fluence_\text{geo}$, where $\alpha$ is the geomagnetic angle, has been parametrised in terms of $r$, $\dmax$ and the density at shower maximum.
    Then the geomagnetic fluence relates to the $\vxB$ fluence as
    \begin{equation}\label{eq:geo_vxb}
        \fluence_\text{geo} = \frac{\fluence_\vxB}{\left(1 + \frac{\cos\phi}{|\sin\alpha|}\sqrt{a_\text{ce}}\right)^2},
    \end{equation}
    where $\sin\phi$ takes into account the azimuth dependence of the \ce emission.
    
    Finally, the measured energy fluence $\fluence_\vxB$ can be modelled as
    \begin{equation}
        \fluence_\vxB = f_\text{GS}(r | E_\text{geo}, \dmax, \vec x) \left(1 + \frac{\cos\phi}{|\sin\alpha|}\sqrt{a_\text{ce}(r|\dmax)}\right)^2,
    \end{equation}
    which is used to find the observables in a $\chi^2$-minimisation fit. This fit has 4 degrees of freedom: $E_\text{geo}$ and $\dmax$ and the position $\vec x = (x, y)$.
    The full fit is only attempted when there are 4 or more stations with signal. If there are only 3 stations that have a signal, the position is not fitted.

\section{Expected performance}
    To estimate the resolution and aperture of the \rd achievable with this reconstruction method, an extensive simulation study has been made on the basis of 7972 showers simulated with CORSIKA/CoREAS~\cite{Schlueter2022_1000149113}.
    The detector simulation for the \rd is relatively straightforward. To obtain the signal at the voltage level it suffices to apply the response equation for the electric field (eq.~\ref{eq:E_to_V}).
    For a realistic simulation, the voltages are converted to \adc counts and measured noise is added. The resulting time traces are then analysed with the methods described above.
    To find the resolution the reconstructed quantities were compared to the ground truth of the simulations.
    From this it was determined, that the fit does not have sufficient constraining power on the distance to shower maximum ($\dmax$).
    For direction and energy this method can achieve an angular resolution of $\sim 0.2\deg$ and an energy resolution of $\sim5\%$.
    This shows that the standard reconstruction method for the \rd is mature and ready for the operation of the detector. 
    For a look at first data, see~\cite{bjarni_this_icrc}.

\bibliographystyle{JHEP}
\footnotesize
\bibliography{bibliography}

\clearpage

\section*{The Pierre Auger Collaboration}

{\footnotesize\setlength{\baselineskip}{10pt}
\noindent
\begin{wrapfigure}[11]{l}{0.12\linewidth}
\vspace{-4pt}
\includegraphics[width=0.98\linewidth]{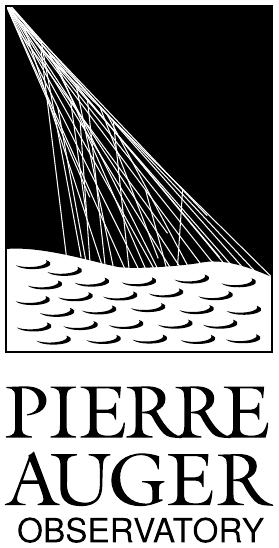}
\end{wrapfigure}
\begin{sloppypar}\noindent
A.~Abdul Halim$^{13}$,
P.~Abreu$^{70}$,
M.~Aglietta$^{53,51}$,
I.~Allekotte$^{1}$,
K.~Almeida Cheminant$^{78,77}$,
A.~Almela$^{7,12}$,
R.~Aloisio$^{44,45}$,
J.~Alvarez-Mu\~niz$^{76}$,
A.~Ambrosone$^{44}$,
J.~Ammerman Yebra$^{76}$
% ...

\end{sloppypar}
\begin{center}
\end{center}

\vspace{1ex}
\begin{description}[labelsep=0.2em,align=right,labelwidth=0.7em,labelindent=0em,leftmargin=2em,noitemsep,before={\renewcommand\makelabel[1]{##1 }}]
\item[$^{1}$] Centro At\'omico Bariloche and Instituto Balseiro (CNEA-UNCuyo-CONICET), San Carlos de Bariloche, Argentina
\item[$^{2}$] Departamento de F\'\i{}sica and Departamento de Ciencias de la Atm\'osfera y los Oc\'eanos, FCEyN, Universidad de Buenos Aires and CONICET, Buenos Aires, Argentina
\item[$^{3}$] IFLP, Universidad Nacional de La Plata and CONICET, La Plata, Argentina
\item[$^{4}$] Instituto de Astronom\'\i{}a y F\'\i{}sica del Espacio (IAFE, CONICET-UBA), Buenos Aires, Argentina
\item[$^{5}$] Instituto de F\'\i{}sica de Rosario (IFIR) -- CONICET/U.N.R.\ and Facultad de Ciencias Bioqu\'\i{}micas y Farmac\'euticas U.N.R., Rosario, Argentina
\item[$^{6}$] Instituto de Tecnolog\'\i{}as en Detecci\'on y Astropart\'\i{}culas (CNEA, CONICET, UNSAM), and Universidad Tecnol\'ogica Nacional -- Facultad Regional Mendoza (CONICET/CNEA), Mendoza, Argentina
\item[$^{7}$] Instituto de Tecnolog\'\i{}as en Detecci\'on y Astropart\'\i{}culas (CNEA, CONICET, UNSAM), Buenos Aires, Argentina
\item[$^{8}$] International Center of Advanced Studies and Instituto de Ciencias F\'\i{}sicas, ECyT-UNSAM and CONICET, Campus Miguelete -- San Mart\'\i{}n, Buenos Aires, Argentina
\item[$^{9}$] Laboratorio Atm\'osfera -- Departamento de Investigaciones en L\'aseres y sus Aplicaciones -- UNIDEF (CITEDEF-CONICET), Argentina
\item[$^{10}$] Observatorio Pierre Auger, Malarg\"ue, Argentina
\item[$^{11}$] Observatorio Pierre Auger and Comisi\'on Nacional de Energ\'\i{}a At\'omica, Malarg\"ue, Argentina
\item[$^{12}$] Universidad Tecnol\'ogica Nacional -- Facultad Regional Buenos Aires, Buenos Aires, Argentina
\item[$^{13}$] University of Adelaide, Adelaide, S.A., Australia
\item[$^{14}$] Universit\'e Libre de Bruxelles (ULB), Brussels, Belgium
\item[$^{15}$] Vrije Universiteit Brussels, Brussels, Belgium
\item[$^{16}$] Centro Brasileiro de Pesquisas Fisicas, Rio de Janeiro, RJ, Brazil
\item[$^{17}$] Centro Federal de Educa\c{c}\~ao Tecnol\'ogica Celso Suckow da Fonseca, Petropolis, Brazil
\item[$^{18}$] Instituto Federal de Educa\c{c}\~ao, Ci\^encia e Tecnologia do Rio de Janeiro (IFRJ), Brazil
\item[$^{19}$] Universidade de S\~ao Paulo, Escola de Engenharia de Lorena, Lorena, SP, Brazil
\item[$^{20}$] Universidade de S\~ao Paulo, Instituto de F\'\i{}sica de S\~ao Carlos, S\~ao Carlos, SP, Brazil
\item[$^{21}$] Universidade de S\~ao Paulo, Instituto de F\'\i{}sica, S\~ao Paulo, SP, Brazil
\item[$^{22}$] Universidade Estadual de Campinas (UNICAMP), IFGW, Campinas, SP, Brazil
\item[$^{23}$] Universidade Estadual de Feira de Santana, Feira de Santana, Brazil
\item[$^{24}$] Universidade Federal de Campina Grande, Centro de Ciencias e Tecnologia, Campina Grande, Brazil
\item[$^{25}$] Universidade Federal do ABC, Santo Andr\'e, SP, Brazil
\item[$^{26}$] Universidade Federal do Paran\'a, Setor Palotina, Palotina, Brazil
\item[$^{27}$] Universidade Federal do Rio de Janeiro, Instituto de F\'\i{}sica, Rio de Janeiro, RJ, Brazil
\item[$^{28}$] Universidad de Medell\'\i{}n, Medell\'\i{}n, Colombia
\item[$^{29}$] Universidad Industrial de Santander, Bucaramanga, Colombia
\item[$^{30}$] Charles University, Faculty of Mathematics and Physics, Institute of Particle and Nuclear Physics, Prague, Czech Republic
\item[$^{31}$] Institute of Physics of the Czech Academy of Sciences, Prague, Czech Republic
\item[$^{32}$] Palacky University, Olomouc, Czech Republic
\item[$^{33}$] CNRS/IN2P3, IJCLab, Universit\'e Paris-Saclay, Orsay, France
\item[$^{34}$] Laboratoire de Physique Nucl\'eaire et de Hautes Energies (LPNHE), Sorbonne Universit\'e, Universit\'e de Paris, CNRS-IN2P3, Paris, France
\item[$^{35}$] Univ.\ Grenoble Alpes, CNRS, Grenoble Institute of Engineering Univ.\ Grenoble Alpes, LPSC-IN2P3, 38000 Grenoble, France
\item[$^{36}$] Universit\'e Paris-Saclay, CNRS/IN2P3, IJCLab, Orsay, France
\item[$^{37}$] Bergische Universit\"at Wuppertal, Department of Physics, Wuppertal, Germany
\item[$^{38}$] Karlsruhe Institute of Technology (KIT), Institute for Experimental Particle Physics, Karlsruhe, Germany
\item[$^{39}$] Karlsruhe Institute of Technology (KIT), Institut f\"ur Prozessdatenverarbeitung und Elektronik, Karlsruhe, Germany
\item[$^{40}$] Karlsruhe Institute of Technology (KIT), Institute for Astroparticle Physics, Karlsruhe, Germany
\item[$^{41}$] RWTH Aachen University, III.\ Physikalisches Institut A, Aachen, Germany
\item[$^{42}$] Universit\"at Hamburg, II.\ Institut f\"ur Theoretische Physik, Hamburg, Germany
\item[$^{43}$] Universit\"at Siegen, Department Physik -- Experimentelle Teilchenphysik, Siegen, Germany
\item[$^{44}$] Gran Sasso Science Institute, L'Aquila, Italy
\item[$^{45}$] INFN Laboratori Nazionali del Gran Sasso, Assergi (L'Aquila), Italy
\item[$^{46}$] INFN, Sezione di Catania, Catania, Italy
\item[$^{47}$] INFN, Sezione di Lecce, Lecce, Italy
\item[$^{48}$] INFN, Sezione di Milano, Milano, Italy
\item[$^{49}$] INFN, Sezione di Napoli, Napoli, Italy
\item[$^{50}$] INFN, Sezione di Roma ``Tor Vergata'', Roma, Italy
\item[$^{51}$] INFN, Sezione di Torino, Torino, Italy
\item[$^{52}$] Istituto di Astrofisica Spaziale e Fisica Cosmica di Palermo (INAF), Palermo, Italy
\item[$^{53}$] Osservatorio Astrofisico di Torino (INAF), Torino, Italy
\item[$^{54}$] Politecnico di Milano, Dipartimento di Scienze e Tecnologie Aerospaziali , Milano, Italy
\item[$^{55}$] Universit\`a del Salento, Dipartimento di Matematica e Fisica ``E.\ De Giorgi'', Lecce, Italy
\item[$^{56}$] Universit\`a dell'Aquila, Dipartimento di Scienze Fisiche e Chimiche, L'Aquila, Italy
\item[$^{57}$] Universit\`a di Catania, Dipartimento di Fisica e Astronomia ``Ettore Majorana``, Catania, Italy
\item[$^{58}$] Universit\`a di Milano, Dipartimento di Fisica, Milano, Italy
\item[$^{59}$] Universit\`a di Napoli ``Federico II'', Dipartimento di Fisica ``Ettore Pancini'', Napoli, Italy
\item[$^{60}$] Universit\`a di Palermo, Dipartimento di Fisica e Chimica ''E.\ Segr\`e'', Palermo, Italy
\item[$^{61}$] Universit\`a di Roma ``Tor Vergata'', Dipartimento di Fisica, Roma, Italy
\item[$^{62}$] Universit\`a Torino, Dipartimento di Fisica, Torino, Italy
\item[$^{63}$] Benem\'erita Universidad Aut\'onoma de Puebla, Puebla, M\'exico
\item[$^{64}$] Unidad Profesional Interdisciplinaria en Ingenier\'\i{}a y Tecnolog\'\i{}as Avanzadas del Instituto Polit\'ecnico Nacional (UPIITA-IPN), M\'exico, D.F., M\'exico
\item[$^{65}$] Universidad Aut\'onoma de Chiapas, Tuxtla Guti\'errez, Chiapas, M\'exico
\item[$^{66}$] Universidad Michoacana de San Nicol\'as de Hidalgo, Morelia, Michoac\'an, M\'exico
\item[$^{67}$] Universidad Nacional Aut\'onoma de M\'exico, M\'exico, D.F., M\'exico
\item[$^{68}$] Institute of Nuclear Physics PAN, Krakow, Poland
\item[$^{69}$] University of \L{}\'od\'z, Faculty of High-Energy Astrophysics,\L{}\'od\'z, Poland
\item[$^{70}$] Laborat\'orio de Instrumenta\c{c}\~ao e F\'\i{}sica Experimental de Part\'\i{}culas -- LIP and Instituto Superior T\'ecnico -- IST, Universidade de Lisboa -- UL, Lisboa, Portugal
\item[$^{71}$] ``Horia Hulubei'' National Institute for Physics and Nuclear Engineering, Bucharest-Magurele, Romania
\item[$^{72}$] Institute of Space Science, Bucharest-Magurele, Romania
\item[$^{73}$] Center for Astrophysics and Cosmology (CAC), University of Nova Gorica, Nova Gorica, Slovenia
\item[$^{74}$] Experimental Particle Physics Department, J.\ Stefan Institute, Ljubljana, Slovenia
\item[$^{75}$] Universidad de Granada and C.A.F.P.E., Granada, Spain
\item[$^{76}$] Instituto Galego de F\'\i{}sica de Altas Enerx\'\i{}as (IGFAE), Universidade de Santiago de Compostela, Santiago de Compostela, Spain
\item[$^{77}$] IMAPP, Radboud University Nijmegen, Nijmegen, The Netherlands
\item[$^{78}$] Nationaal Instituut voor Kernfysica en Hoge Energie Fysica (NIKHEF), Science Park, Amsterdam, The Netherlands
\item[$^{79}$] Stichting Astronomisch Onderzoek in Nederland (ASTRON), Dwingeloo, The Netherlands
\item[$^{80}$] Universiteit van Amsterdam, Faculty of Science, Amsterdam, The Netherlands
\item[$^{81}$] Case Western Reserve University, Cleveland, OH, USA
\item[$^{82}$] Colorado School of Mines, Golden, CO, USA
\item[$^{83}$] Department of Physics and Astronomy, Lehman College, City University of New York, Bronx, NY, USA
\item[$^{84}$] Michigan Technological University, Houghton, MI, USA
\item[$^{85}$] New York University, New York, NY, USA
\item[$^{86}$] University of Chicago, Enrico Fermi Institute, Chicago, IL, USA
\item[$^{87}$] University of Delaware, Department of Physics and Astronomy, Bartol Research Institute, Newark, DE, USA
\item[] -----
\item[$^{a}$] Max-Planck-Institut f\"ur Radioastronomie, Bonn, Germany
\item[$^{b}$] also at Kapteyn Institute, University of Groningen, Groningen, The Netherlands
\item[$^{c}$] School of Physics and Astronomy, University of Leeds, Leeds, United Kingdom
\item[$^{d}$] Fermi National Accelerator Laboratory, Fermilab, Batavia, IL, USA
\item[$^{e}$] Pennsylvania State University, University Park, PA, USA
\item[$^{f}$] Colorado State University, Fort Collins, CO, USA
\item[$^{g}$] Louisiana State University, Baton Rouge, LA, USA
\item[$^{h}$] now at Graduate School of Science, Osaka Metropolitan University, Osaka, Japan
\item[$^{i}$] Institut universitaire de France (IUF), France
\item[$^{j}$] now at Technische Universit\"at Dortmund and Ruhr-Universit\"at Bochum, Dortmund and Bochum, Germany
\end{description}

\section*{Acknowledgments}

\begin{sloppypar}
The successful installation, commissioning, and operation of the Pierre
Auger Observatory would not have been possible without the strong
commitment and effort from the technical and administrative staff in
Malarg\"ue. We are very grateful to the following agencies and
organizations for financial support:
\end{sloppypar}

\begin{sloppypar}
Argentina -- Comisi\'on Nacional de Energ\'\i{}a At\'omica; Agencia Nacional de
Promoci\'on Cient\'\i{}fica y Tecnol\'ogica (ANPCyT); Consejo Nacional de
Investigaciones Cient\'\i{}ficas y T\'ecnicas (CONICET); Gobierno de la
Provincia de Mendoza; Municipalidad de Malarg\"ue; NDM Holdings and Valle
Las Le\~nas; in gratitude for their continuing cooperation over land
access; Australia -- the Australian Research Council; Belgium -- Fonds
de la Recherche Scientifique (FNRS); Research Foundation Flanders (FWO),
Marie Curie Action of the European Union Grant No.~101107047; Brazil --
Conselho Nacional de Desenvolvimento Cient\'\i{}fico e Tecnol\'ogico (CNPq);
Financiadora de Estudos e Projetos (FINEP); Funda\c{c}\~ao de Amparo \`a
Pesquisa do Estado de Rio de Janeiro (FAPERJ); S\~ao Paulo Research
Foundation (FAPESP) Grants No.~2019/10151-2, No.~2010/07359-6 and
No.~1999/05404-3; Minist\'erio da Ci\^encia, Tecnologia, Inova\c{c}\~oes e
Comunica\c{c}\~oes (MCTIC); Czech Republic -- GACR 24-13049S, CAS LQ100102401,
MEYS LM2023032, CZ.02.1.01/0.0/0.0/16{\textunderscore}013/0001402,
CZ.02.1.01/0.0/0.0/18{\textunderscore}046/0016010 and
CZ.02.1.01/0.0/0.0/17{\textunderscore}049/0008422 and CZ.02.01.01/00/22{\textunderscore}008/0004632;
France -- Centre de Calcul IN2P3/CNRS; Centre National de la Recherche
Scientifique (CNRS); Conseil R\'egional Ile-de-France; D\'epartement
Physique Nucl\'eaire et Corpusculaire (PNC-IN2P3/CNRS); D\'epartement
Sciences de l'Univers (SDU-INSU/CNRS); Institut Lagrange de Paris (ILP)
Grant No.~LABEX ANR-10-LABX-63 within the Investissements d'Avenir
Programme Grant No.~ANR-11-IDEX-0004-02; Germany -- Bundesministerium
f\"ur Bildung und Forschung (BMBF); Deutsche Forschungsgemeinschaft (DFG);
Finanzministerium Baden-W\"urttemberg; Helmholtz Alliance for
Astroparticle Physics (HAP); Helmholtz-Gemeinschaft Deutscher
Forschungszentren (HGF); Ministerium f\"ur Kultur und Wissenschaft des
Landes Nordrhein-Westfalen; Ministerium f\"ur Wissenschaft, Forschung und
Kunst des Landes Baden-W\"urttemberg; Italy -- Istituto Nazionale di
Fisica Nucleare (INFN); Istituto Nazionale di Astrofisica (INAF);
Ministero dell'Universit\`a e della Ricerca (MUR); CETEMPS Center of
Excellence; Ministero degli Affari Esteri (MAE), ICSC Centro Nazionale
di Ricerca in High Performance Computing, Big Data and Quantum
Computing, funded by European Union NextGenerationEU, reference code
CN{\textunderscore}00000013; M\'exico -- Consejo Nacional de Ciencia y Tecnolog\'\i{}a
(CONACYT) No.~167733; Universidad Nacional Aut\'onoma de M\'exico (UNAM);
PAPIIT DGAPA-UNAM; The Netherlands -- Ministry of Education, Culture and
Science; Netherlands Organisation for Scientific Research (NWO); Dutch
national e-infrastructure with the support of SURF Cooperative; Poland
-- Ministry of Education and Science, grants No.~DIR/WK/2018/11 and
2022/WK/12; National Science Centre, grants No.~2016/22/M/ST9/00198,
2016/23/B/ST9/01635, 2020/39/B/ST9/01398, and 2022/45/B/ST9/02163;
Portugal -- Portuguese national funds and FEDER funds within Programa
Operacional Factores de Competitividade through Funda\c{c}\~ao para a Ci\^encia
e a Tecnologia (COMPETE); Romania -- Ministry of Research, Innovation
and Digitization, CNCS-UEFISCDI, contract no.~30N/2023 under Romanian
National Core Program LAPLAS VII, grant no.~PN 23 21 01 02 and project
number PN-III-P1-1.1-TE-2021-0924/TE57/2022, within PNCDI III; Slovenia
-- Slovenian Research Agency, grants P1-0031, P1-0385, I0-0033, N1-0111;
Spain -- Ministerio de Ciencia e Innovaci\'on/Agencia Estatal de
Investigaci\'on (PID2019-105544GB-I00, PID2022-140510NB-I00 and
RYC2019-027017-I), Xunta de Galicia (CIGUS Network of Research Centers,
Consolidaci\'on 2021 GRC GI-2033, ED431C-2021/22 and ED431F-2022/15),
Junta de Andaluc\'\i{}a (SOMM17/6104/UGR and P18-FR-4314), and the European
Union (Marie Sklodowska-Curie 101065027 and ERDF); USA -- Department of
Energy, Contracts No.~DE-AC02-07CH11359, No.~DE-FR02-04ER41300,
No.~DE-FG02-99ER41107 and No.~DE-SC0011689; National Science Foundation,
Grant No.~0450696, and NSF-2013199; The Grainger Foundation; Marie
Curie-IRSES/EPLANET; European Particle Physics Latin American Network;
and UNESCO.
\end{sloppypar}

}

\end{document}